\documentclass[aps,prd,twocolumn,superscriptaddress,nofootinbib,floatfix]{revtex4-2}

\usepackage{amsmath,amssymb}
\usepackage{graphicx}
\usepackage{booktabs}
\usepackage{xcolor}
\usepackage[colorlinks=true,linkcolor=blue!60!black,citecolor=blue!60!black,urlcolor=blue!60!black]{hyperref}

\graphicspath{{figures/}}

\newcommand{\phenjax}{\texttt{phenjax}}
\newcommand{\phenomxpy}{\texttt{phenomxpy}}
\newcommand{\phentax}{\texttt{phentax}}
\newcommand{\jax}{\textsc{jax}}
\newcommand{\Msun}{M_\odot}

\begin{document}

\title{\phenjax: the IMRPhenomT waveform family in JAX for accelerated analyses on CPUs and GPUs}

\author{Andrea Begnoni}
\email{andrea.begnoni@uib.es}
\affiliation{IAC3, Universitat de les Illes Balears, Cra. de Valldemossa km 7.5, 07122 Palma, Spain}
\date{\today}

\begin{abstract} 
We present \phenjax, a \jax{} implementation of the precessing time-domain IMRPhenomT family of models for the gravitational wave signal emitted in compact binary coalescence. The code keeps the structure of its reference Python implementation, \phenomxpy, and reproduces, on average, its polarizations to a few $\times 10^{-12}$ of the peak strain for all the models, aligned-spin and precessing. 
The waveform models now take advantage of the \jax{} construction: compiled once per array shape, vectorized over events (batching), differentiable and executed on CPUs or GPUs with the same source code. Different additional features address the cost of long signals and of heterogeneous populations in a batch. On one NVIDIA H100, for the precessing TPHM model on LISA-mass populations, \phenjax{} is 3 to 6 times faster than \phenomxpy{} when both codes evaluate one event at a time and 20 to 30 times faster when \phenjax{} evaluates the events in batches of 8. We estimate that, for waveform generation, one GPU does the work of 1300 to 2700 cores of a current CPU node at 25 to 45 times less energy per waveform. As a backend of a global-fit LISA massive black hole pipeline, performed on the Sangria dataset, \phenjax{} recovers the same posteriors as the other waveform backends at a lower computational cost. In a comparison with \phentax{}, the wall time was 13 minutes instead of 21.
\end{abstract}

\maketitle

\section{Introduction}
\label{sec:intro}

Gravitational wave (GW) parameter estimation of compact binary coalescences is a computationally intensive process and its principal cost is the evaluation of the waveform model, which connects the source parameters to the GW strain measured by the detector. 
The implementation of most waveform models has been successfully built for a decade on models written in C inside the LIGO Algorithm Library (LAL) \cite{lalsuite}. Its design was intended for its scope in the LIGO-Virgo-KAGRA collaboration; in fact, it was envisioned to be used by a sampler calling one waveform at a time, running on CPUs, where each waveform call is relatively cheap because the signals of stellar-mass binaries in current ground-based detectors contain relatively few samples. Two developments now ask for additional capabilities that stress this original design. First, the next generation of detectors brings longer signals and larger catalogues: the Einstein Telescope (ET) will observe binary neutron stars for hours in band and detect $\mathcal{O}(10^5)$ events per year \cite{Maggiore:2019uih, Branchesi:2023mws}, while the Laser Interferometer Space Antenna (LISA) will observe massive black hole binaries (MBHBs) for days to months \cite{LISA:2024hlh}, so that a single waveform contains up to millions of samples. Second, samplers increasingly evaluate the waveform for many events all at once, on GPUs and, sometimes, together with its derivatives \cite{Wong:2023lgb,Wong:2022xvh,Karnesis:2023ras,Katz:2020hku,Katz:2024oqg,Demasi:2026smc}. 

Array frameworks such as \jax{} \cite{jax2018github} offer a different trade-off. The model is written once in Python, with a syntax close to NumPy and the same source is compiled for CPUs and GPUs, vectorized over events and differentiated automatically. This lowers the cost of implementing and modifying a model.
The price is a set of constraints imposed by the compiler which drive several design choices described in Sec.~\ref{sec:design}.
Several codes have followed this route for the Phenom models: for example, GWFAST contains frequency-domain models reimplemented in \jax{} \cite{Iacovelli:2022bbs,Iacovelli:2022mbg} while \texttt{ripple} ports several frequency-domain IMRPhenom models to \jax{} and uses them with the gradient-based sampler of \texttt{jim} \cite{Edwards:2023sak, Wong:2023lgb}. Recently, \phentax{} was released, a \jax{} version of the time-domain IMRPhenomT and IMRPhenomTHM models \cite{Santini:2026phentax}.
GWJulia follows the same approach for frequency-domain models, but it reimplements them in Julia \cite{Begnoni:2025oyd}. 

The IMRPhenomT family \cite{Estelles:2020twz,Estelles:2020osj,Estelles:2021gvs} models the dominant and subdominant modes of aligned-spin binaries in closed form and twists them up for precession \cite{Schmidt:2012rh} with Euler angles obtained either from the next-to-next-to-leading-order (NNLO) post-Newtonian precession equations \cite{Bohe:2012mr}, from the multiple-scale analysis (MSA)~\cite{Chatziioannou:2017tdw} or by integrating the spin evolution numerically, using ODEs. The family already has a Python implementation, \phenomxpy{} \cite{Garcia-Quiros:2025usi}, accelerated with \texttt{numba} on CPUs and \texttt{CuPy} on GPUs \cite{numba,cupy}, which parallelizes over the time samples of one event. 
Regarding precessing models with higher modes, i.e., PhenomTPHM, \phenomxpy{} is the only Python implementation to date in the time domain, while \texttt{ripple} has the only precessing frequency-domain implementation, IMRPhenomXPHM \cite{Pratten:2020ceb,Edwards:2023sak,ripple_github}.

In this work, we present \phenjax, a \jax{} port of \phenomxpy{} for the models T, THM, TP and TPHM with the NNLO and MSA angles. The guiding principle of the port is that neither the physics nor the methods used can change: the code keeps the familiar structure of \phenomxpy, which is derived from LAL, its fits, its mode content and its conventions and every model is compared with its version of \phenomxpy{} on a population of events. What \phenjax{} adds is the advantages of \jax{}, i.e., compilation once per array shape, vectorization over events, gradients with respect to the parameters and CPU or GPU execution with the same source code. On top of it, we provide various variants of the code to maximize performance across different use cases and hardware.

The paper is organized as follows. Section~\ref{sec:design} describes the structure of the code, the bucketing scheme, the GPU variants and the benchmark suite. Section~\ref{sec:results} reports the accuracy with respect to \phenomxpy, the waveform benchmarks on CPUs and GPUs, and the use of \phenjax{} as a backend of a LISA MBHB pipeline on the Sangria dataset of the LISA Data Challenges. Section~\ref{sec:conclusions} summarizes the results and the planned extensions.

\section{Motivations and structure}
\label{sec:design}

\subsection{Structure and interface}
\label{sec:structure}

\phenjax{} mirrors \phenomxpy{}: the same classes (\texttt{pWF}, \texttt{pPhase}, \texttt{pAmp} and the \texttt{IMRPhenomT*} approximants), the same calibration coefficients and the same methods, with the NumPy and \texttt{numba} kernels replaced by \texttt{jax.numpy} and the control flow written so that it traces under \texttt{jit} and \texttt{vmap}. The API is the same as \phenomxpy{}, with the addition of the \jax{}-specific features. The API covers T, THM, TP and TPHM with either precession prescription and Table~\ref{tab:models} lists the models.

\begin{table}[h!]
\caption{Models in the release. The numerical prescription for the precession angles, obtained by integrating the spin evolution, is the planned next step (Sec.~\ref{sec:conclusions}).}
\label{tab:models}
\footnotesize\setlength{\tabcolsep}{3pt}
\begin{ruledtabular}
\begin{tabular}{llll}
Model & Spins & Precession angles & Status \\
\hline
T & aligned & -- & released \\
THM & aligned & -- & released \\
TP / TPHM NNLO & precessing & closed form (PN) & released \\
TP / TPHM MSA & precessing & closed form (MSA) & released \\
TP / TPHM numerical & precessing & spin ODEs & planned \\
\end{tabular}
\end{ruledtabular}
\end{table}

The code, leveraging the vectorization capabilities of \jax{}, evaluates a rectangular batch of waveforms with one compiled call of \texttt{vmap(waveform\_single)}. 
Therefore, every batch must consist of events of the same length and the code is compiled once per batch shape. Recompilations can be costly and these constraints result from \jax{}'s requirement that each function have static array shapes. This is due to XLA (Accelerated Linear Algebra), which is the underlying compiler for \jax{} and automatically optimizes and transforms Python code to run natively on CPUs, GPUs or TPUs. A waveform generator that chooses its own number of time samples from the event parameters, as \phenomxpy{} does, cannot be compiled once. Thus, \phenjax{} offers three ways to set the time grid. The user can pass a time array, which is evaluated as given, as in \phenomxpy{}. The user can also set a length, in which case the code computes the natural length of the event from its parameters and raises an error if the target is shorter than it. Or the code can compute the natural length itself and build the grid, at the cost of a small overhead. When a batch of events is evaluated on one grid, the length of the grid must be at least that of the longest event in the batch. The batched grid builders do not check it: if the rows end at the end of each event, as in \phentax{}, a grid shorter than an event drops its earliest samples, while if the rows start at the beginning of each event, the layout used by the bucketing scheme, the row stops before the merger. The length check is therefore left to the user, who can build the grid from the natural lengths computed by the code. 
To address the cost of long signals, Sec.~\ref{sec:bucketing} describes a bucketing scheme that evaluates events in groups of similar length and Sec.~\ref{sec:variants} describes kernel variants that reduce the per-sample cost. 

\subsection{Bucketing}
\label{sec:bucketing}

\begin{figure}[h!]
\centering
\includegraphics[width=\columnwidth]{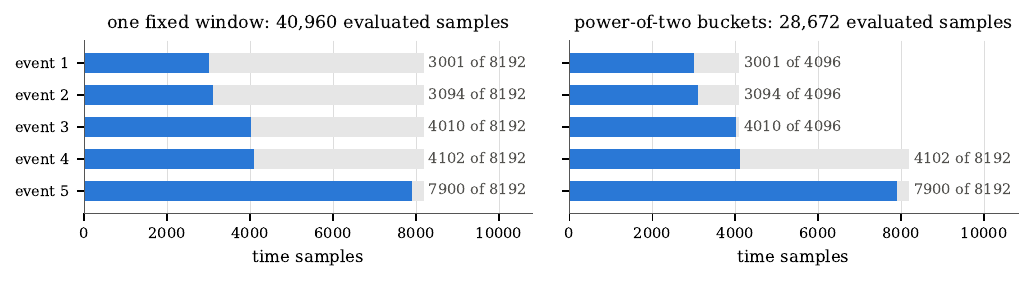}
\caption{Bucketing on five events with various natural lengths (blue). One fixed window (left) extends every event to 8192 samples; power-of-two buckets (right) evaluate the first three events in a $(3, 4096)$ batch and the other two in a $(2, 8192)$ batch. The grey extension is evaluated as a ringdown.}
\label{fig:bucketing}
\end{figure}

As said before, a batch of events must share the same time vector length, due to XLA constraints, but the duration of a signal from a fixed starting frequency scales as $M^{-5/3}$ with the total mass $M$. For a population spanning a decade in mass, setting the length of every event to the longest one evaluates and stores, for the short events, up to $\sim 50$ times more samples than they need. \phenjax{} can precalculate the exact time length (also considering spins), assign each event of natural length $N_i$ to a grid whose length is the power of two above $N_i$, $T_i = 2^{\lceil \log_2 N_i \rceil}$, and evaluate one rectangular batch per bucket. Since $N_i \le T_i < 2 N_i$, no event is extended beyond twice its natural length, whatever the population and there is one compilation per bucket. For an example of the procedure, see Figure~\ref{fig:bucketing}. The advantage of the procedure is a reduced number of samples evaluated and a reduced memory usage; however, it requires a larger number of compilations and an overhead for each event. The balance between the two depends both on the population and on the hardware used; therefore, the user can choose between the use of one fixed window for all events or the bucketing scheme. 

In particular, the saving in evaluated samples does not imply a saving in time on every device and every analysis. In fact, each bucket is a separate launch with a smaller batch and, as shown in Sec.~\ref{sec:batching}, a launch has a fixed cost of a few milliseconds on a GPU. For this reason, bucketing is a clear gain on CPUs and for populations whose lengths spread over many buckets and a smaller or negative one on fast GPUs for populations of similar lengths, where one fixed window in one program is already efficient. The benchmarks of Sec.~\ref{sec:buckets} quantify both cases.

\subsection{GPU kernel variants}
\label{sec:variants}

\begin{figure}[t]
\includegraphics[width=\columnwidth]{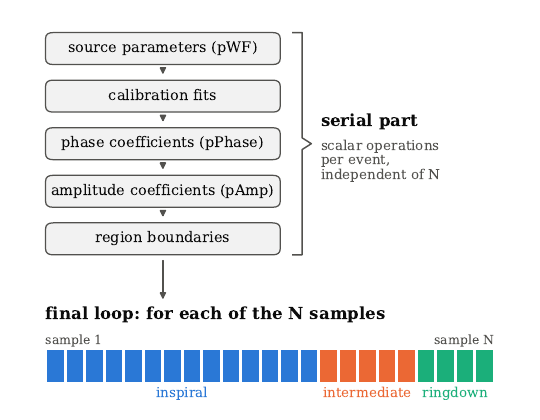}
\caption{Structure of one waveform evaluation. The serial part is a chain of scalar operations per event, from the source parameters to the boundaries of the inspiral, intermediate and ringdown regions, and does not depend on the number of samples $N$. The final loop evaluates every sample; under \texttt{vmap}, the formulas of the three regions are evaluated on every sample and only then selected.}
\label{fig:structure}
\end{figure}

A waveform can be divided into two parts: a long serial series of computations independent of the number of samples and a final loop over samples. Figure~\ref{fig:structure} sketches the two parts. The final step is the most expensive part for long signals (Sec.~\ref{sec:batching} quantifies where the crossover lies); thus, it is critical to optimize it.
The first part is a chain of scalar operations per event and it is parallel over the events only; thus, a batch of eight events gives only eight independent work items. Therefore, when this computation is performed on a GPU, this part provides far too few threads to occupy the device, but its cost can be amortized easily by increasing the batch size.
The second part is compiled into elementwise kernels in which every thread handles its own samples. This part consists of a long series of `if' statements that select the formulas for the inspiral, intermediate and ringdown regions of the waveform for each sample. In the original \phenomxpy{} port, the second part is not efficient for batches in \jax{} for two reasons. Firstly, it fills the number of threads available even on a modern GPU long before the GPU can be fully utilized. The result is that the GPU will not be able to run efficiently and we see unexpected inefficiencies in the performance with large batches, see Sec.~\ref{sec:batching}. Secondly, the use of \jax{} with `if' statements via \texttt{vmap} requires that all the branches are evaluated and only then selected. This is particularly inefficient considering that for the long and expensive signals, most of the time will be spent in the inspiral part, but the code will also evaluate the intermediate part and the ringdown for all samples.

To tackle this inefficiency, we have implemented a set of variants of the final loop over samples. The original idea was to improve the computation on the GPUs, but we have found that the variants are also more efficient on CPUs. The original port will be referred to as the baseline, while the GPU variants are divided into tier~1 and tier~2. Tier~1 variants keep using float64 for all the computations, while tier~2 uses float32 for some quantities, sacrificing precision (an error of at most $2.7\times 10^{-7}$ of the peak strain on the worst sample, with respect to the baseline) for speed.
\begin{itemize}
\item \emph{Mode loop.} In the baseline code, the higher modes are evaluated together as one array with one entry per mode and per sample; this is convenient but has some additional costs.
The mode loop computes the frequency and the phase of the $(2,2)$ mode once, before the loop, and then evaluates the modes one after the other in a single plain Python loop over $(\ell, m)$; every mode, the $(2,2)$ included, reads the same precomputed series and each mode is added to the strain as soon as it is computed. For T and TP, which contain only the $(2,2)$ mode, it evaluates only that mode.
\item \emph{Tail split.} A long signal spends almost all of its samples in the inspiral, but the three-way selection evaluates the inspiral, intermediate and ringdown formulas on every sample. To avoid this, the tail split divides the time axis of each event into an inspiral bulk, where only the inspiral formulas are evaluated and a merger window of $16384$ samples located with a dynamic slice around the peak, where the three-region evaluators run. A runtime check verifies that the split was done correctly: if any bulk sample lies past the earliest region boundary, the batch falls back to the full evaluation, so the result never depends on the guess.
\item \emph{Shared transcendentals.} The final for loop contains many transcendental terms that become computationally expensive when evaluated millions of times. The transcendental terms that are shared by the modes (the powers of the orbital frequency, the trigonometric functions of the phase) are now evaluated once per sample and the mode phasors $e^{-im\psi/2}$ are obtained by complex products of one cosine and one sine.
\end{itemize}
These three steps together define the \emph{region split} variant, which is in tier 1. They are exact up to roundoff; in fact, on every GPU tested, the region split reproduces the polarizations of the baseline \phenjax{} code to a maximum relative error of $2.5\times10^{-12}$ of the peak strain, so its accuracy with respect to \phenomxpy{} is that of the baseline (Sec.~\ref{sec:accuracy}). 
Further variants exist and they are described in the code repository, but they are not used in the benchmarks of this paper.
We noticed that GPUs with slow float64 operations can benefit from using mixed precision for certain quantities; therefore, we implemented a mixed-precision variant in tier 2.
\begin{itemize}
\item \emph{Mixed precision (tier~2).} The quantities that do not accumulate phase (amplitude polynomials, phasor powers, quaternion products, Wigner rows and the mode sum) are evaluated in float32, while the frequency, the phase and every trigonometric function of a large argument stay in float64. This is the mixed-precision form of the region split: its maximum relative error is $2.7\times10^{-7}$ and its mismatch with the baseline code, computed with a flat noise spectrum, is below $10^{-13}$, well inside the tolerance of any current analysis. However, since it alters the roundoff of the model, it is kept separate.
\end{itemize}

The variants are selected by the user while the default remains the baseline. The region split is the recommended setting for long signals on any GPU and the mixed-precision form for GPUs with slow float64 operations. The variants currently cover all the released models. 

\subsection{Benchmarks shipped with the code}
\label{sec:suite}

Whether batching, bucketing and the variants pay depends on the length of the signals, the batch size the memory allows, the GPU model and, in a sampler, on how the sampler builds and launches its program. There are more combinations than a user can reasonably explore. For this reason, the repository ships the benchmark suite with the code, together with the complete records of the campaigns reported here: the seeded population, the timing, memory, power and accuracy of every configuration, on multiple hardware platforms, i.e., two CPU models and five GPU models, as JSON records and as Markdown reports written to be read by a user or by a Large Language Model (LLM) assistant. We suggest that users who need to choose a configuration for a given analysis and hardware read the reports closest to their case, or let an LLM digest them, perform a few benchmarks, compare the results and only then launch the full analysis. 
Note that automatic differentiation has not been reviewed yet; the code allows for it, but it should be used with care.
\phenjax{} is developed for the analyses of the LISA Distributed Data Processing Center (DDPC) and of ET, where long signals, large populations and batched evaluation on GPUs are important needs. It is released under the MIT license and distributed on request: the code, the GPU variants, the benchmark suite and the records of the campaigns of this paper are available to everyone from the author. 

\section{Results}
\label{sec:results}

\subsection{Accuracy with respect to \phenomxpy}
\label{sec:accuracy}

Every release of \phenjax{} is compared with \phenomxpy{} on a seeded population of 100 events per model, with total mass in $[30, 120]\,\Msun$, spin magnitudes up to $0.95$ and starting frequencies in $[20, 30]$~Hz, each code building its own time grid from the parameters of the event. The comparison covers separately the calibration fits, the individual modes, the Euler angles, the time grids (which agree to $5\times10^{-16}$) and the polarizations. Table~\ref{tab:accuracy} reports the polarization errors normalized to the peak strain of each event.
For each event, the error at every time sample is $|h^{\rm jax} - h^{\rm xpy}| / \max |h^{\rm xpy}|$ and the mean averages it over the samples of each event and then over the 100 events, while the maximum is the single worst sample of the whole campaign.
\begin{table}[t]
\caption{Polarization errors of \phenjax{} with respect to \phenomxpy, $|h^{\rm jax} - h^{\rm xpy}| / \max |h^{\rm xpy}|$, over 100 events per model: mean over samples and events and worst sample of the 100-event campaign.}
\label{tab:accuracy}
\scriptsize\setlength{\tabcolsep}{2.5pt}
\begin{ruledtabular}
\begin{tabular}{lcccc}
Model & $h_+$ mean & $h_+$ max & $h_\times$ mean & $h_\times$ max \\
\hline
T         & $1.7\times10^{-12}$ & $8.1\times10^{-11}$ & $1.7\times10^{-12}$ & $9.0\times10^{-11}$ \\
THM       & $1.5\times10^{-12}$ & $1.1\times10^{-10}$ & $1.5\times10^{-12}$ & $1.2\times10^{-10}$ \\
TP NNLO   & $1.7\times10^{-12}$ & $6.2\times10^{-11}$ & $1.7\times10^{-12}$ & $6.0\times10^{-11}$ \\
TP MSA    & $2.8\times10^{-12}$ & $1.5\times10^{-7}$  & $2.9\times10^{-12}$ & $3.3\times10^{-7}$ \\
TPHM NNLO & $1.5\times10^{-12}$ & $6.4\times10^{-11}$ & $1.5\times10^{-12}$ & $6.9\times10^{-11}$ \\
TPHM MSA  & $2.5\times10^{-12}$ & $1.4\times10^{-7}$  & $2.6\times10^{-12}$ & $2.9\times10^{-7}$ \\
\end{tabular}
\end{ruledtabular}
\end{table}

The aligned-spin and NNLO models agree with the reference at the $10^{-10}$ level on the worst sample. The MSA models agree at the same level on average and reach $3\times10^{-7}$ on one event of the campaign, where the error enters through the Euler angles ($6\times10^{-8}$ on the angles of that event) and is amplified by the twisting up.
The noise-free mismatch between the two implementations stays below $10^{-11}$ on every event tested,
far below any observational threshold. Since the fits, the modes and the methods are those of \phenomxpy, the accuracy of \phenjax{} against numerical relativity is expected to be similar to that of the IMRPhenomT family \cite{Estelles:2020twz,Estelles:2020osj,Estelles:2021gvs}.

\subsection{Waveform benchmarks on CPUs and GPUs}
\label{sec:benchmarks}

\subsubsection{Setup}

Unless stated otherwise, the reference campaign ran on MareNostrum5 at the Barcelona Supercomputing Center: a CPU node with two Xeon Platinum 8480+ (112 cores) and one H100 (64~GB) of an exclusively reserved accelerated node. Each run draws 64 events with mass ratio uniform in $[1, 20]$, spin magnitudes uniform in $[0, 0.99]$ with isotropic directions and distances uniform in $[0.1, 10]$~Gpc. The total mass, which, with the sampling time, sets the length of the signal, is analyzed for different scenarios. \phenjax{} evaluates them on one fixed window set by the longest event, either one event at a time (one launch per event) or in batches of 8 (8 launches per pass), \phenomxpy{} one at a time with its \texttt{CuPy} backend on the GPU, and \phentax{} (T and THM only) in the same batches of 8.
Times are warm medians, with compilation recorded separately and excluded and the GPU power is sampled during the timed passes. All the GPU numbers of this paper and the comparison with \phenomxpy{} and \phentax{} on the CPU node were taken in one software stack: \jax{} 0.11.2, CuPy 14.2 and \texttt{numba} 0.67.
The kernel variants were also measured on the H200, B200 and L40S GPUs of the TalaIA cluster of the University of the Balearic Islands and on a Blackwell workstation GPU. In the following, we call the H100, H200 and B200 data-centre GPUs (high performance) and the L40S and the Blackwell workstation GPU workstation GPUs (low performance) and name the model only where a number refers to a specific one.

\subsubsection{Launch cost, per-sample cost and the gain of batching}
\label{sec:batching}

\begin{figure}[t]
\includegraphics[width=\columnwidth]{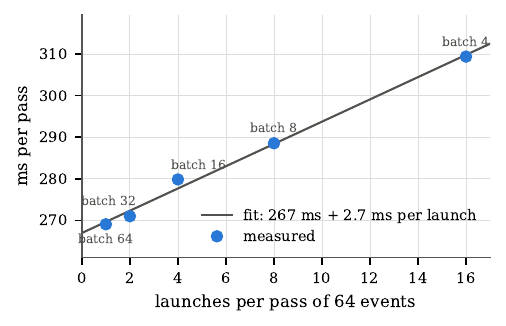}
\caption{Time of a pass of 64 TPHM-NNLO events at $10^6\,\Msun$ on the H100 against the number of launches of the compiled program, from batch 64 (one launch) to batch 4 (sixteen launches), with the two-term fit of Eq.~\eqref{eq:twoterm}.}
\label{fig:launch}
\end{figure}

As mentioned before, batching means evaluating several events in one call of the compiled program, so that the fixed cost of the launch of the program is paid once per batch instead of once per event and the GPU receives enough independent work to fill its threads. The question is how much of the time of a pass, i.e., the total time to compute a given number of waveforms, it can remove. 
Figure~\ref{fig:launch} shows the time of a pass of 64 TPHM-NNLO events at $10^6\,\Msun$ (2.08M samples per waveform, a long signal) on the H100 as the batch size goes from 4 to 64, i.e., from 16 launches of the compiled program to one. We used the baseline code; the picture changes slightly with the other GPU variants introduced in Sec.~\ref{sec:variants}. A linear fit against the number of launches gives
\begin{equation}
t_{\rm pass} = 267~\text{ms} + 2.7~\text{ms} \times (\text{launches}),
\label{eq:twoterm}
\end{equation}
with residuals below 2~ms. The first term, the y-intercept, is the cost of the 133M samples of the pass (the 64 waveforms of 2.08M samples), i.e., 2.0~ns per sample for this model and this baseline and does not depend on how the events are grouped. The second, the one containing the slope, is the fixed cost of one execution of the program (a few hundred fused kernels, each with its launch latency and tail). Batching cuts the launch cost exactly and it is independent of the length of the signal.
A shorter signal would have a smaller y-intercept, i.e., a smaller sample term, but the inclination of the curve, 2.7~ms, would be the same, so the absolute gain of batching is the same, but proportionally larger. This stays true until the GPU is filled by the samples of the batch; for short signals, batching also affects the per-sample cost as we will see in Fig.~\ref{fig:sweep} below.
The device memory grows linearly from 2 to 31~GiB and it is not saturated. The crossover between the launch-dominated and the sample-dominated regimes is near $10^6$ samples per waveform on the GPU for the baseline, above which we call the event a long event for our purposes.

Equation~\eqref{eq:twoterm} also indicates where we need to focus our attention: the cost per sample takes most of the time in the pass and it does not change with batching.
The reason is that the baseline requires too many threads for the GPU to keep resident, yet it runs at a few per cent of both the float64 peak and the memory bandwidth of the device: each sample is a long dependent chain of float64 operations with many transcendental functions.
This issue was tackled using the kernel variants of Sec.~\ref{sec:variants}, which reduce the per-sample cost and improve the throughput. This can be seen in Figure~\ref{fig:sweep}, where we show the time per event against the batch size, on the H100, for a short ($10^7\,\Msun$, 144k samples) and a long ($10^5\,\Msun$, 4.04M samples) population, with the baseline and two variants. For the long signals, the baseline (the black curve) is flat from 4 events on, at 2.8~ms per event for THM and 8.3~ms for TPHM NNLO, whereas the variants keep gaining up to 64 events, where the variants are about 5 times faster. The behaviour of the black curve is the one expected from Eq.~\eqref{eq:twoterm}, where the per-sample term dominates and batching does not help.
The variants have cut the per-sample term so far that the per-launch floor of about 3.5~ms matters again and the GPU is filled only from 32 to 64 events. For the short signals, where the launch cost dominates, the variants are significantly faster only at 128 events and above, reaching improvements of about a factor of 2 at 1024 events. This way, the variants recover the behaviour one expects from batching: the time per event keeps falling with the batch until the device is full, for long and short signals alike. Note that the variants are not always faster: for example, for short signals and small batches, the baseline is slightly faster.

\begin{figure*}[t]
\includegraphics[width=\textwidth]{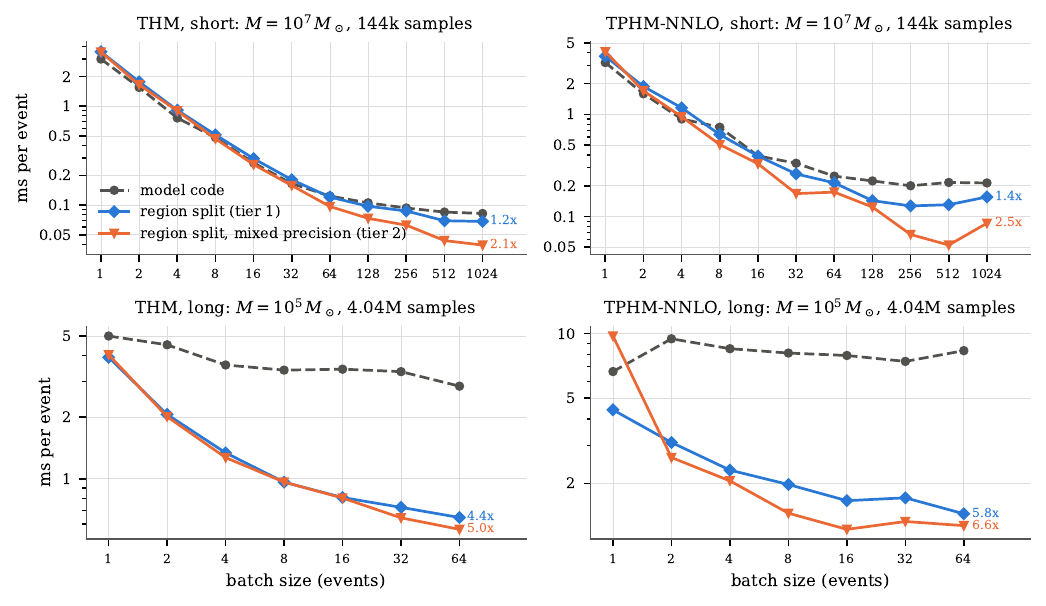}
\caption{Time per event against the batch size on a GPU for THM (left) and TPHM NNLO (right), for a short (top) and a long (bottom) LISA population, with the baseline of the model code, the tier~1 region split and its tier~2 mixed-precision form. The labels give the speed-up over the baseline at the largest batch.}
\label{fig:sweep}
\end{figure*}

\subsubsection{Kernel variants on five GPU models}
\label{sec:fivegpus}

\begin{figure*}[t]
\includegraphics[width=\textwidth]{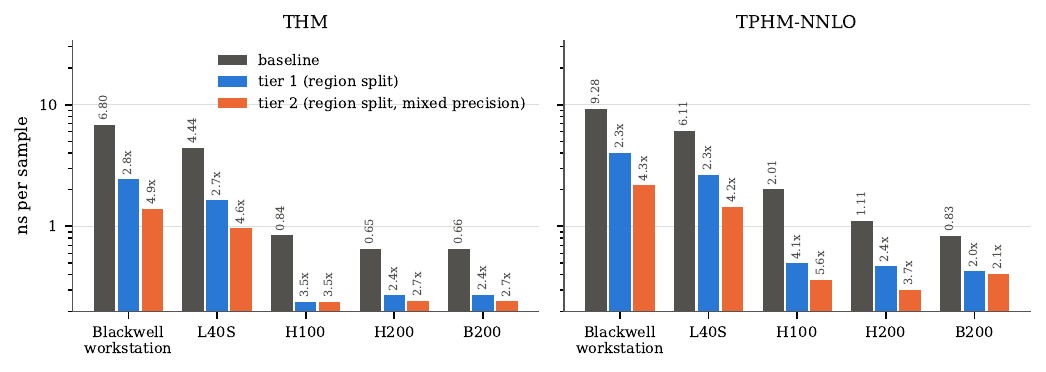}
\caption{Per-sample cost of THM and TPHM NNLO for the long LISA population ($10^5\,\Msun$, 8 events of 4.04M samples) on five GPU models: the baseline, the tier~1 region split and its tier~2 mixed-precision form, with the speed-up over the baseline of the same GPU.}
\label{fig:variants}
\end{figure*}

Different users may have access to different hardware configurations and even within GPU families, the performance can vary significantly. Figure~\ref{fig:variants} reports the per-sample cost of the baseline and of the two variants for the long population (8 events of 4.04M samples) on five GPU models. The three high-performance GPUs have similar performance and end at the same 0.24~ns per sample for THM, while mixed precision adds little, and at 0.30 to 0.41~ns per sample for TPHM NNLO with mixed precision (0.43 to 0.50~ns with tier~1), i.e., 1.3 to 1.7 times the cost of THM. On the two workstation GPUs, whose float64 pipeline is slower, the mixed precision is a significant improvement. The per-sample cost of the baseline spans a factor of 10 across the five GPUs. 
One thing to note is that different variants within the same tier have different levels of improvement on different GPUs; thus, we also share this information in the reports shipped with the code, so that users can choose the best variant for their GPU.

\subsubsection{Bucketing against the fixed window}
\label{sec:buckets}

Bucketing trades evaluated samples for launches and compilations, so whether it pays depends on the device and on how spread the lengths are. On a CPU, samples are expensive and launches cheap: for a broad ground-based population ($M$ in $[5, 200]\,\Msun$ at 20~Hz), where the buckets cut the samples of a pass by a factor of 16, THM runs 10 times faster with buckets on the CPU node. On a fast GPU, the balance is different, because every bucket is a separate, smaller launch with its own fixed cost and its own compilation; thus, the compilation time grows significantly.
We measured four LISA populations of 256 events on a data-centre GPU (H200) and a workstation GPU (L40S), with fixed and with sampled masses, with the baseline and with the bucket forms of the variants. Three rules of thumb summarize the results. Bucketing is the default on CPUs and for batched populations spread over a wide range of lengths, where it gains on every device, up to 4 times on the L40S. On a data-centre GPU with the variants, one fixed window is the first thing to try, since it was the fastest option on the H200 in almost all cases. On a workstation GPU the two gains combine: for the populations with sampled masses on the L40S, the bucket form of the tier~2 variant is faster than both the bucketed baseline and the tier~2 variant on one fixed window, by 1.1 to 1.6 times over the better of the two. The reports shipped with the code (Sec.~\ref{sec:suite}) give the numbers for every case.

\subsubsection{CPU cores against GPUs}
\label{sec:node}

\begin{figure*}[t]
\includegraphics[width=\textwidth]{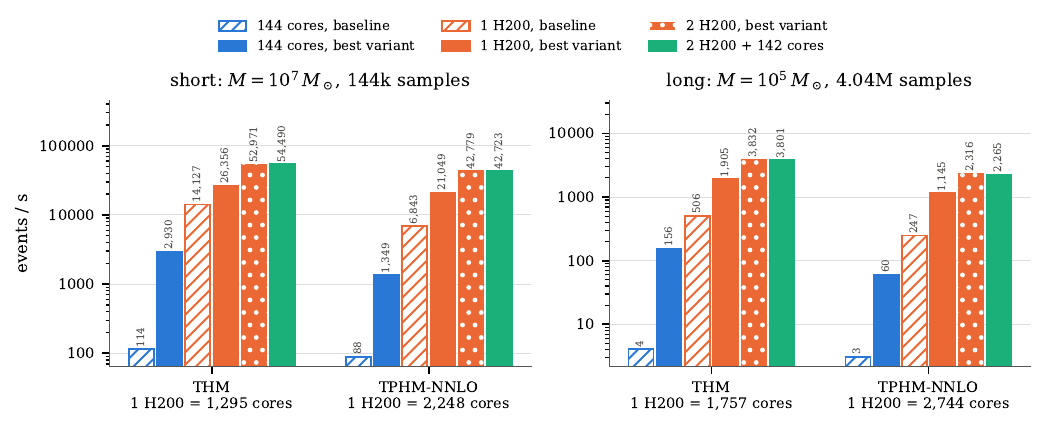}
\caption{Throughput of the 144 cores and of the H200 GPUs for the short (left) and long (right) LISA population, with the baseline (hatched) and the best variant (full) and the sum of two GPUs and 142 cores measured side by side. The axis labels give the number of cores one H200 is worth with the best code on each device. The last bar of each group is the throughput of the whole node, measured with the 142 cores and the two GPUs running at the same time.}
\label{fig:node}
\end{figure*}

The most frequent question about a GPU port is how many CPU cores one GPU replaces. We measured it on one CPU node with two AMD EPYC 9565 (144 cores) and two H200: on the CPU side, 144 single-threaded \jax{} processes pinned one per core, each repeating the same batch for 30~s, while on the GPU side, one process per H200 with the batch size that gives the highest throughput for that signal on the GPU (within its memory: 1024 short or 64 long events). Every configuration was run with the baseline and with the variants, for the short ($10^7\,\Msun$, 144k samples) and the long ($10^5\,\Msun$, 4.04M samples) LISA population, each evaluated on a single fixed window of that length, i.e., without bucketing. Figure~\ref{fig:node} shows the results and three facts stand out. First, the baseline is very slow on CPUs under XLA, 9 to 12~$\mu$s per sample per core, and the variant of tier~1 is roughly 20 to 40 times faster per core; therefore, the variants are not only a GPU optimization. Second, in the same campaign we also ran one process with 144 XLA threads (not shown in the figure) and found that it is 3 to 9 times less efficient than 144 single-threaded processes. XLA on CPUs parallelizes inside each operation and pays only for large arrays, whereas independent waveforms are embarrassingly parallel; therefore, CPU users should run one single-threaded process per core. Third, with the best code on each device, one data-centre GPU is worth 1,300 to 2,700 of these cores, i.e., roughly 10 to 20 of these 144-core nodes. Running the whole node at once, the two GPUs, each fed by one dedicated core, and the remaining 142 cores, gives 2 to 5\% less than the sum of the separate measurements. In fact, the cores keep their rate, while the GPUs lose 3 to 5\% when the host is fully loaded and the cores add 3 to 6\% to what the two GPUs do in the same run, so that the whole node delivers between 2\% less and 3\% more than the two GPUs alone, which makes the CPUs of little use in this campaign.

Energy efficiency is another significant achievement of the port. Counting the CPUs at their thermal design power (800~W for the two sockets) and measuring the GPUs with \texttt{nvidia-smi}, the CPU node spends 0.33~J per short THM event and 15.8~J per long TPHM-NNLO event with the variants, against 0.013 and 0.36~J on one H200: 25 to 45 times less energy per waveform on the GPU. Similar energy savings are expected in comparison to \phenomxpy{} on CPUs and GPUs.

\subsubsection{Comparison with \phenomxpy{} and \phentax{} on LISA populations}
\label{sec:lisa}

\begin{figure*}[t]
\includegraphics[width=\textwidth]{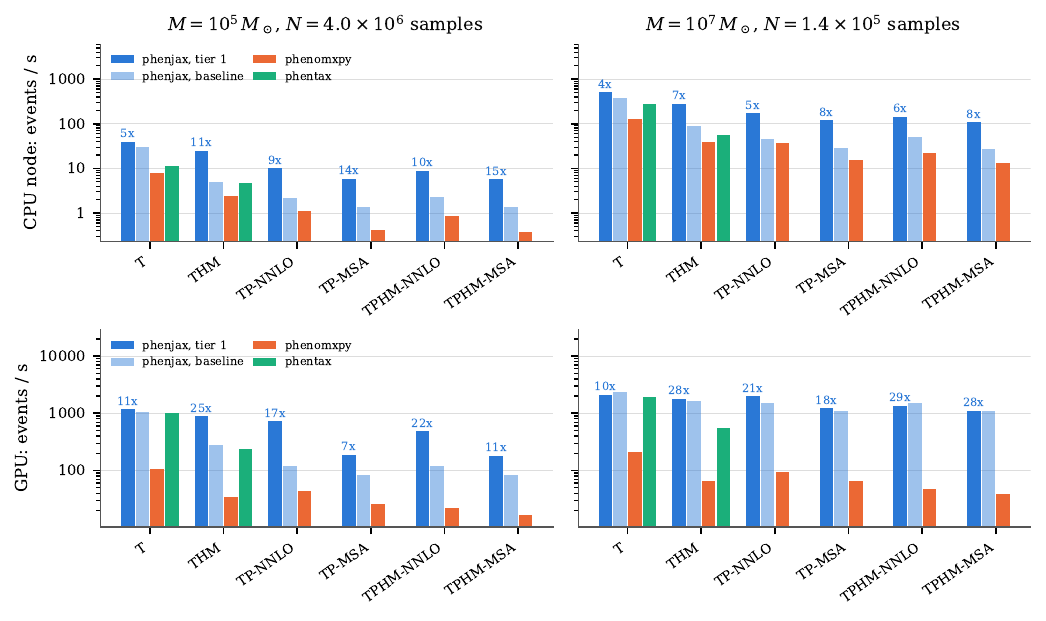}
\caption{Events per second on the CPU node (top) and on the GPU (bottom) for the LISA populations of $10^5$ and $10^7\,\Msun$: 64 events on one fixed window, warm timings; $N$ is the number of samples per waveform of the window. \phenjax{} runs in batches of 8 events with the baseline and with the tier~1 variant, \phenomxpy{} runs one event at a time, \phentax{}, which provides T and THM only, runs in the same batches of 8. The label above each tier~1 bar is its speed-up over \phenomxpy{} on the same device.}
\label{fig:lisa}
\end{figure*}

We can now benchmark \phenjax{} against \phenomxpy{} and \phentax{}, i.e., the other codes for IMRPhenomT models in Python.
Figure~\ref{fig:lisa} shows the throughput for two LISA populations, $M = 10^5$ and $10^7\,\Msun$ with $f_{\rm min} = 1.2\times10^{-3}$ and $10^{-4}$~Hz and $\Delta t = 1$ and 10~s, whose fixed windows hold 4.0M and 0.14M samples per waveform and Table~\ref{tab:h100} gives the pass times of THM and TPHM NNLO on a GPU with and without batching. Two numbers summarize the precessing model. One event at a time, \phenjax{} with the tier~1 variant is 3 to 6 times faster than \phenomxpy{}: both codes then parallelize over the samples of one event and the gain is that of one fused XLA program over a sequence of \texttt{CuPy} kernels, both on the GPU. In batches of 8 events in tier~1, the long population takes 0.13~s per pass against 2.89~s, about 20 times faster and the short one 0.047~s against 1.35~s, about 30 times.
The baseline in the same batches of 8 events is 5 to 30 times faster than \phenomxpy{}, but on long signals almost none of this gain comes from batching, as Sec.~\ref{sec:batching} showed. The gain of batching itself grows with the mass, i.e., with shorter windows, because \phenomxpy{} parallelizes over the samples of one event and cannot fill the GPU with a short one, while a batch of \phenjax{} does. For the aligned models \phentax{}, also an XLA program, is as fast as \phenjax{} for T and 3 to 4 times slower for THM in the same batches, due to the \emph{mode loop} of Sec.~\ref{sec:variants}. On the CPU node (top row of the figure) the baseline gains 1.3 to 3.9 times over \phenomxpy{}, since \phenomxpy{} is already vectorized with \texttt{numba} over the time axis and the tier~1 variant gains 4 to 15 times in batches of 8. These CPU timings use one process with the XLA thread pool, as a sampler would.
Section~\ref{sec:node} shows that one single-threaded process per core is the efficient CPU layout. The cost of precession within \phenjax{} can also be read from the figure. For the long signals, where the per-sample work dominates, the baseline evaluates TPHM NNLO 2.3 times slower than THM and TP NNLO 8.6 times slower than T,
while the variant brings the two ratios to 1.8 and 1.6. For the short signals, where the launch cost dominates, a precessing model costs up to 1.6 times its aligned counterpart with either code.
As we showed before, these results are heavily dependent on the batch size, so the best way to compare the two is to run the two codes on the same hardware using the user-dependent setup. We did not come across any case where \phenomxpy{} was faster than \phenjax{} on the same hardware and the gain of \phenjax{} over \phenomxpy{} is larger for long signals, which are the most expensive to evaluate.

\begin{table}[t]
\caption{Warm time of a pass of 64 events on the GPU (seconds): \phenjax{} with the baseline and with the tier~1 variant, one event at a time (1) and in batches of 8 (8), the tier~2 variant in batches of 8, \phenomxpy{} (xpy) with its \texttt{CuPy} backend one event at a time, and the faster \phentax{} (tax) execution in batches of 8, for the $10^5\,\Msun$ (4.04M samples per waveform) and $10^7\,\Msun$ (0.14M) populations; TPHM is TPHM NNLO.}
\label{tab:h100}
\scriptsize\setlength{\tabcolsep}{2pt}
\begin{ruledtabular}
\begin{tabular}{llrrrrrrr}
 & & \multicolumn{2}{c}{baseline} & \multicolumn{2}{c}{tier~1} & tier~2 & & \\
$M$ & Model & 1 & 8 & 1 & 8 & 8 & xpy & tax \\
\hline
$10^5\,\Msun$ & THM  & 0.516 & 0.233 & 0.466 & 0.072 & 0.073 & 1.84 & 0.269 \\
$10^5\,\Msun$ & TPHM & 0.595 & 0.534 & 0.493 & 0.133 & 0.101 & 2.89 & -- \\
$10^7\,\Msun$ & THM  & 0.399 & 0.039 & 0.450 & 0.035 & 0.037 & 0.976 & 0.115 \\
$10^7\,\Msun$ & TPHM & 0.434 & 0.043 & 0.460 & 0.047 & 0.040 & 1.35 & -- \\
\end{tabular}
\end{ruledtabular}
\end{table}

\subsection{LISA analysis of Sangria data with GeeMoo}
\label{sec:geemoo}

The LISA use case that motivated the long-signal work is the parameter estimation of MBHBs within a global-fit pipeline \cite{Littenberg:2023xpl,Strub:2024kbe,Katz:2024oqg,Deng:2025wgk}, e.g., GeeMoo \cite{mbhb_fastlike} or Erebor \cite{Katz:2024oqg}. In the case of GeeMoo, a \jax{} likelihood evaluates the waveform, the LISA response on the six links, the TDI combinations $X$, $Y$, $Z$ and the log-likelihood in one compiled function, sampled with the sampler \texttt{jexplore} with all chains evaluated at once through \texttt{vmap}. \phenjax{} can be used as a backend of the pipeline next to \phentax{} and \phenomxpy{}.

\begin{figure}[t]
\includegraphics[width=\columnwidth]{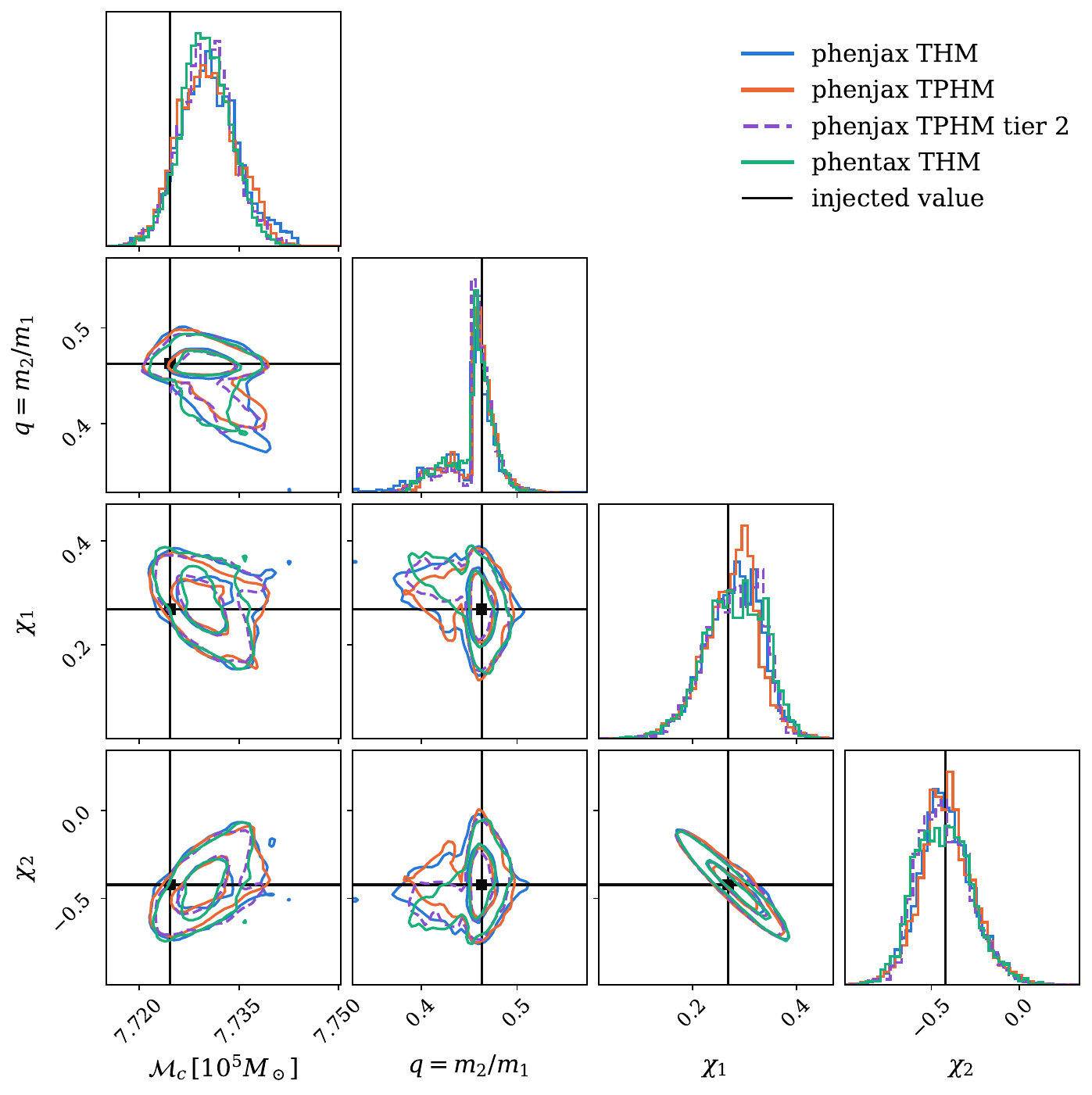}
\caption{Posteriors of the intrinsic parameters of the Sangria MBHB~4 (distance rescaled) injected in the Sangria data without its MBHBs, from GeeMoo runs with the \phenjax{} THM, \phenjax{} TPHM (aligned spins), \phenjax{} TPHM tier~2 and \phentax{} THM backends, same data, priors and seed. The contours enclose 39 and 86\% of the probability and the lines mark the injected values.}
\label{fig:corner}
\end{figure}

We analysed an MBHB injected into the public training data of Sangria, the second LISA Data Challenge \cite{Baghi:2022ucj}.
We inject the Sangria MBHB~4 with its catalogue parameters\footnote{Parameters: $m_1=1.32\times10^6\,M_\odot$, $m_2=6.12\times10^5\,M_\odot$ in the detector frame, $\chi_{1z}=0.27$, $\chi_{2z}=-0.42$, luminosity distance 36.8~Gpc; sky position, orientation and coalescence time from the Sangria catalogue.}, except the luminosity distance, which we rescale to obtain a signal-to-noise ratio of 150, with modes $(2,2)$, $(2,1)$, $(3,3)$, $(4,4)$, analysed from 7 days before to 1 day after merger at 5~s (138k samples) in the band $4\times10^{-5}$ to $4\times10^{-3}$~Hz. Note that all MBHB events in Sangria are non-precessing.
The sampler ran 24 walkers at 5 temperatures for 10,000 iterations, $1.2\times10^6$ likelihood evaluations per run, with the same data, priors and seed for \phenjax{} THM, \phenjax{} TPHM (with zero precessing spins), \phenjax{} TPHM with the tier~2 variant and \phentax{} THM. Note that the times for TPHM take into account the extra cost of precession, i.e., the Euler angles are evaluated and the modes are twisted, but the precessing spins are zero.
The LISA response is the \jax{} implementation of the pipeline.
Figure~\ref{fig:corner} shows that the four runs recover the injection with the same posteriors.

Table~\ref{tab:geemoo} reports where the time goes in the likelihood evaluation. With \phenjax{} the waveform is a third to a half of the likelihood step, while the LISA response and the inner product are the rest. \phenjax{} THM evaluates the waveform 2.0 times faster than \phentax{} THM and the full likelihood 1.4 times faster. We expect that for more refined orbits, like the ones in Mojito \cite{mojito}, the LISA response and the inner product will become the dominant factors in the likelihood evaluation. Similarly, the kernel variants, which gain at most 1.2 times on the waveform here (the window is short, 138k samples and the chains form one batch of 120), do not change the likelihood time significantly. A production run of 10,000 iterations on one H200 takes 13 minutes with \phenjax{} THM, 15 with TPHM, 21 with TPHM tier~2 and 21.1 with \phentax{} THM, against many hours on the 144 cores of the same node.
At this point, the waveform is no longer the bottleneck of a GeeMoo-like analysis while the full LISA response, the likelihood inner product and the compilation of the sampler are.

\begin{table}[t]
\caption{Time of one evaluation of the batch of 120 chains in GeeMoo on one H200, in milliseconds: the waveform alone, the full log-likelihood in float64 and the likelihood with its inner product in mixed precision, for the baseline and the tier~2 variant of \phenjax{} and for \phentax{}.}
\label{tab:geemoo}
\footnotesize\setlength{\tabcolsep}{3pt}
\begin{ruledtabular}
\begin{tabular}{lrrr}
 & waveform & likelihood & mixed \\
\hline
THM, baseline & 10.5 & 27.3 & 21.3 \\
THM, tier 2 & 9.3 & 26.8 & 20.7 \\
TPHM NNLO, baseline & 16.5 & 33.6 & 27.4 \\
TPHM NNLO, tier 2 & 13.8 & 29.9 & 23.4 \\
\phentax{} THM & 20.7 & 37.5 & 32.2 \\
\end{tabular}
\end{ruledtabular}
\end{table}

\section{Conclusions}
\label{sec:conclusions}
We have presented \phenjax, a \jax{} port of the IMRPhenomT family that keeps the structure and the physics of \phenomxpy{} and reproduces its polarizations on average to $10^{-12}$ of the peak strain.
The port adds just-in-time compilation, vectorization over events and GPU execution, a bucketing scheme for populations of heterogeneous lengths and a set of kernel variants that evaluate the long events more efficiently. On one GPU, for TPHM on LISA populations, \phenjax{} is 3 to 6 times faster than \phenomxpy{} one event at a time and 20 to 30 times faster in batches of 8, considering the variants. One GPU does the work of 1,300 to 2,700 cores of a current CPU node at 25 to 45 times less energy per waveform.

The benchmarks also delimit what batching can do. The time of a pass is a per-launch term of a few milliseconds plus a per-sample term and batching only removes launches: it pays in full below $10^6$ samples per waveform and little above, unless the per-sample term is reduced, which is what the variants do. Bucketing trades evaluated samples for launches and compilation and is therefore a clear gain on CPUs and for populations spread over many octaves and a case-by-case one on fast GPUs. As a backend of GeeMoo on Sangria data, \phenjax{} recovers the same posteriors as \phentax{} in 60\% of the time on one H200 and the waveform is a third to a half of the likelihood: the LISA response and the inner product are now the bottleneck of this class of analyses.

The gradients of the waveforms are being tested; they can be used but are not yet reviewed. Moreover, the numerical prescription for the precession angles, which integrates the spin evolution and is the most accurate option of the family for generic spins, is the next model we plan to add, with the integration written so that it batches and differentiates like the rest of the code. 

There is ongoing work to include multibanding, which is a technique that reduces the number of samples of a waveform by evaluating it on a coarse grid and interpolating it on a finer one, with the goal of reducing the per-sample cost of long signals. 
Finally, a natural follow-up of this work is to port the IMRPhenomX family, which is the current state of the art of the phenomenological models.

\begin{acknowledgments}
The author thanks Sascha Husa, under whose guidance this work was started and carried out. The author would also like to thank Cecilio Garc{\'\i}a-Quir{\'o}s, H{\'e}ctor Estell{\'e}s, Antoni Ramos-Buades and Jorge Valencia. The author is also grateful to the developers of \phenomxpy{} and \phentax{}, since this code took inspiration from their work and from the lessons learned from them.
This work was supported by the Universitat de les Illes Balears (UIB) with funds from the Programa de Foment de la Recerca i la Innovació de la UIB 2024-2026 (supported by the yearly plan of the Tourist Stay Tax ITS2023-086); the Spanish Agencia Estatal de Investigación grants PID2022-138626NB-I00, RED2024-153978-E and RED2024-153735-E, funded by MICIU/AEI/10.13039/501100011033 and the ERDF/EU; and the Comunitat Autònoma de les Illes Balears 
through the Conselleria d'Educació i Universitats with funds from the European Union -- European Regional Development Fund (ERDF) (SINCO2022/18146 -- Plataforma HiTech-IAC3-BIO).
The author gratefully acknowledges the supercomputing resources provided by the Balearic Islands Centre for Supercomputing and AI (BSAI). Significant portions of the analysis were run in the GPP and ACC partitions of the TalaIA supercomputer.
The author thankfully acknowledges the computer resources at MareNostrum and the technical support provided by the Barcelona Supercomputing Center (BSC) through grant No.~AECT-2026-1-0034 from the Red Española de Supercomputación (RES).

\end{acknowledgments}

\section*{Data availability}
The code and the data that support the findings of this article are not yet publicly available. The \phenjax{} code, released under the MIT license and the benchmark records of Sec.~\ref{sec:suite} are available upon request.

\section*{AI declaration}
The author acknowledges the use of Claude and ChatGPT for the generation of text and code. The author takes full responsibility for the content of the article.

\bibliography{refs}

\end{document}